\documentclass[useAMS,usenatbib]{mn2e}
\usepackage{graphicx}
\usepackage{latexsym}
\usepackage{url}
\usepackage{psfig}

\title[FIR colours of transition discs]{Far-infrared signatures and inner hole sizes of protoplanetary discs undergoing inside-out dust dispersal}

\author[Ercolano, Koepferl, Owen, Robitaille]{Barbara Ercolano$^{1,2}$\thanks{E-mail: ercolano@usm.lmu.de (BE)}, Christine Koepferl$^{3}$,  James Owen$^{4,5}$, Thomas Robitaille$^{3}$\\
$^{1}$Universit\"ats-Sternwarte M\"unchen, Scheinerstr. 1, 81679 M\"unchen, Germany\\
$^{2}$Excellence Cluster Origin and Structure of the Universe,
Boltzmannstr.2, 85748 Garching bei M\"unchen, Germany\\
$^{3}$Max Planck Institute for Astronomy, K\"onigstuhl 17, D-69117 Heidelberg, Germany\\
$^4$Institute for Advanced Study, Einstein Drive, Princeton, NJ 08540, USA\\
$^5$Hubble Fellow}

\begin{document}

\pagerange{\pageref{firstpage}--\pageref{lastpage}} \pubyear{2011}

\maketitle

\label{firstpage}

\def\nat{Nature}
\def\mnras{MNRAS}
\def\apj{ApJ}
\def\aap{A\&A}
\def\apjl{ApJL}
\def\apjs{ApJS}
\def\bain{BAIN}
\def\araa{ARA\&A}
\def\pasp{PASP}
\def\aj{AJ}
\def\pasj{PASJ}
\def\ga{\sim}

\begin{abstract}

By means of radiative transfer simulation we study the evolution of the far-infrared colours of protoplanetary discs undergoing inside-out dispersal, often referred to as transition discs. We show that a brightening of the mid and far-infrared emission from these objects is a natural consequence of the removal of the inner disc. Our results can fully explain recent observations of transition discs in the Chamaleon and Lupus star forming regions from the Herschel Gould Belt Survey, which show a higher median for the 70$\mu$m (Herschel PACS 1) band of known transition objects compared with primordial discs. Our theoretical results hence support the suggestion that the 70$\mu$m band may be a powerful diagnostic for the identification of transition discs from photometry data, provided that the inner hole is larger than tens of AU, depending on spectral type. Furthermore we show that a comparison of photometry in the K , 12$\mu$m and 70$\mu$m bands to model tracks can provide a rough, but quick estimate of the inner hole size of these objects, provided their inclination is below $\sim$85 degrees and the inner hole size is again larger than tens of AU.

\end{abstract}

\begin{keywords}
protoplanetary discs 

\end{keywords}

\section{Introduction}

The relevance of protoplanetary disc evolution and dispersal for planet formation models is often remarked in the literature. It is widely accepted that the evolution of the surface density and of the accretion properties of protoplanetary discs are dominated for the largest part of their lives by viscous draining (e.g. Hartmann et al., 1998). In recent years, however, it has also become clear that the final dispersal of these objects require a further, more violent process able to truncate discs' lives much earlier and more abruptly that what would be expected from viscous draining alone (e.g. Kenyon \& Hartmann, 1995; Luhman et al. 2010). Indeed mounting evidence exists to show that disc dispersal happens quickly and proceeds from the inside out (e.g. Ercolano, Clarke \& Hall, 2011; Koepferl et al. 2013) via the formation of so-called transition discs, i.e. discs that have an evacuated inner cavity in dust (or a inner region which is severely depleted in optical depth). 

While the large majority of transition discs (TDs) are identified from their infrared spectral energy distributions (e.g. Strom et al 1989; Skrutskie et al 1990; Calvet et al. 2005; Espaillat et al. 2010), an increasing number of them have also been targets of spatially resolved observations (Andrews et al. 2011), which have revealed interesting features in the geometrical distribution of material, including inner disc asymmetries (Perez et al. 2014) and spiral arms (Christiaens et al. 2014). See Espaillat et al. (2014) for a recent review of observations of TDs. While upcoming ALMA observations may change this picture soon, up to now, the spatially resolved TD observations s are still a minority, and in general they are mainly discs around more massive stars  (Herbig Ae). Herbig Ae stars may actually follow a different disc dispersal path than discs around solar-type stars. As an example, X-ray photoevaporation, which is proposed to be one of the dominant disc dispersal mechanisms (Ercolano et al. 2008, 2009; Owen et al. 2010, 2011, 2012) is unlikely to be efficient for Herbig Ae stars, which generally have a much lower X-ray to bolometric luminosity ratios compared to their lower masses counterparts. Thus the study of TDs around Herbig Ae stars may not be relevant to understanding dispersal around solar type stars.  Hence TDs statistics obtained from infrared surveys of young solar-type stars still bear the most relevance as they provide the most stringent tests for theoretical models of disc dispersal and planet-disc interactions. 

Indeed, understanding the formation and evolution of transition discs (TDs) is the aim of a number of theoretical programs, which seek to match a number of observational constraints on TD statistics (e.g. frequency of transition discs, inner hole distributions, accretion rates etc.). If TDs are objects caught in the act of disc dispersal, they may hold the key to understanding this process in more detail, which influences both the planet formation process and the subsequent evolution via migration (e.g. Alexander \& Pascucci 2012; Ercolano \& Rosotti 2015). However, several different mechanisms may produce cavities or gaps in protoplanetary discs, including photoevaporation from extreme and far ultraviolet and X-ray radiation from the central object (e.g. Alexander et al. 2006, Gorti \& Hollenbach, 2009, Ercolano et al. 2009, Owen et al. 2010), dust coagulation (e.g. Dullemond \& Dominik 2005) and dynamical clearing by giant planets (e.g. Armitage \& Hansen 1999; Rice et al. 2003). These processes are all at play in shaping the evolution of a protoplanetary disc and the question of which combination of those may finally trigger the formation of a cavity or a gap in a given set of discs may be more relevant than seeking a single answer for the formation of all TDs. 

In fact recent observations have shown that TDs do not represent a homogeneous set of objects, and this strongly undermines the concept of  all discs with a cavity or a hole having a common origin. Some TDs with large cavities, for example, still show evidence for vigorous accretion, implying the presence of significant amounts of gas withis the dust cavity, which is problematic to explain by photoevaporation alone (e.g. Rosotti et al. 2013). These objects, which also happen to be very bright at mm wavelengths, were found to be unlikely drawn from the same underlying distribution as 'typical' TDs (Owen \& Clarke 2012). 

In this paper we do not seek to provide an answer to the ambitious question of cavity/gap formation in TDs, instead we concentrate on the observational appearance and evolution of an inside-out growing cavity on the SED of discs in the far infrared (FIR). The motivation for this was given by the recent Herschel observations of young stars in the Chamaleon I and II regions with data from the Herschel Gould Belt survey (Andr\'e et al. 2010) presented by Ribas et al. (2013, from now on R13) . They compared the SED slopes between the 12$\mu$m (WISE 3, W3) and 70$\mu$m (PACS 1, P1), i.e. $\alpha_{W3-P1}$, against slopes between the K  band and W3 ($\alpha_{K-W3}$) and found a clear separation between the Class II objects and the TDs in the sample due to the different shape in their SEDs. All TDs in their sample have $\alpha_{W3-P1} > 0$. Similar results were found by applying the same method to a sample in Lupus (Bustamante et al. 2015). Based on their results they propose that the Herschel P1 (70 $\mu$m) band is a robust tool for the identification of TDs, which is free from systematics that may affect other identification methods. R13 further propose that the brightening in the FIR SEDs of TDs may indicate a different evolutionary path compared to ClassII sources. Keane at al. (2014), on the other hand, compared the SEDs of TDs from different star-forming regions to those of Class II objects in the Taurus molecular cloud and do not find a systematic brightening in the 63$\mu$m continuum for the TDs observed by Herschel. Based on their analysis of the [OI] 63$\mu$m line, which appears to be systematically weaker in the TD sample, they suggest structural differences in the outer disc of TDs or lower gas-to-dust ratios. 

We show in this paper instead that a brightening in the FIR flux of some TDs is a natural radiative transfer consequence of the presence and further enlargement of the area of the inner dust cavity. We note that, while all objects undergoing inside-out-dispersal, seen at low to moderate inclinations (less than approximately 85$^o$), will at some point become brighter in the FIR,  not all {\it observed} TDs are expected to show this brightening as this depends on the hole size at the time of the observations, which controls the temperature at the inner radius for a given spectral type. Furthermore we present a set of colour-colour diagrams, including inside-out clearing model tracks in the FIR that may help with future identification of TDs from photometry data and provide a fast first order estimate for the size of the disc inner hole.

This paper is structured as follows. Section 2 presents a brief summary of our methods, our results are then presented and discussed in Section 3. Finally, our conclusions and a brief summary are given in Section~4.

\section{Modelling Approach}

We use the grids of models presented in Koepferl et al. (2013) to illustrate the colour evolution of discs undergoing inside-out dispersal in the FIR colour-colour plane. Koepferl et al. (2013) considered different modes of disc dispersal for a set of stellar and disc parameters and by means of radiative transfer modeling they traced the evolution of those objects in the infrared colour-colour plane (K-[8] versus K-[24]). We refer to this paper for details about the modeling, and mention here only a few points that are useful to interpret the results in the present work. 

We present models tracks of protoplanetary discs dispersed from the inside-out as a function of their inner hole radius, which is always given as multiples of the dust sublimation radius for a given spectral type. This is approximated following Whitney et al. (2004) as
\begin{equation}
 R_S = R_*\left( \frac{T_S}{T_*}\right)^{-2.1}
 \end{equation}
 \noindent where $R_*$ and $T_*$ are stellar radius and temperature, listed in Table~1 of Koepferl et al. (2013), and $R_S$ is the dust sublimation temperature, set to a standard 1600~K. Thus for the SEDs shown in Figure 1, for a K4 star, $R_S$ = 0.057~AU. 

We use tracks for different disc geometry parameters spanning the ranges given in Table~2 of Koepferl et al. (2013). More specifically, the models explore three different levels of disc flaring, controlled by the parameter B and three different cases of dust settling controlled by the parameter H$_D$. The disk density distribution is defined by the standard flared accretion disc structure (Shakura \& Sunayaev 1973; Lynden-Bell \& Pringle 1974; Pringle 1981; Bjorkman 1997; Hartmann et al 1998). A detailed description of the geometrical set-up and the input parameters is given in {\bf Section 2.2.3} of Robitaille et al. (2006), where the B and H$_D$ parameters are refereed to as $\beta$ and $z_D$ in the equations.  The mass of the primordial discs is taken to be 1$\%$ of the mass of the central object and it is adjusted accordingly as the inner cavity becomes larger. The outer radius of the disc is 100 AU, and the radius of the inner cavity is varied as a multiple of the sublimation radius $R_{\rm sub}$ up to a maximum of 5000~$R_{\rm sub}$ or 100 AU.

Constant opening angle (B=1.0), flared (B=1.13) and strongly flared (B=1.25) primordial disc models are represented by the black, red and green symbols, respectively, in Figures 2 and 3. Strongly settled (H$_D$=0.1), settled (H$_D$=0.5) and mixed (H$_D$=1.0) primordial models are represented by diamonds, stars and triangles, respectively in Figures 2 and 3. In the aid of clarity inside-out clearing tracks are plotted in Figures 2 and 3 for all flaring geometries, but only for the mixed (H$_D$=1.0) case. All models have constant disc mass within a given spectral range. We extract synthetic fluxes in the K , W3 and P1 bands with the aid of the {\sc FluxCompensator} (Koepferl et al. 2015), which can account for instrumental effects, making the synthetic photometry directly comparable to real photometry. The tool can also thread observed images and will be publicly available in the future.\footnote{For further information see \url{www.mpia.de/~koepferl}.} For the filter convolution we use the transmission curve from W3 (Wright et al. 2010 and 
WISE Handbook\footnote{\url{http://wise2.ipac.caltech.edu/docs/release/prelim/expsup/sec4_3g.html}}),  P1 (PACS Handbook\footnote{\url{https://nhscsci.ipac.caltech.edu/sc/index.php/PACS/FilterCurves}}) and K  (2MASS Handbook\footnote{\url{http://www.ipac.caltech.edu/2mass/releases/allsky/doc/sec6_4a.html}} and Cohen et al. 2003). We excluded objects with an optical disc extinction larger than 10, which applies to objects close to edge-on.

In the following section we present our results both in terms of colours and in terms of slopes in the spectral energy distribution. When probing the evolution in colour-colour space or in the spectral slope usually a variety of combinations are used. Koepferl et al. (2013) presented colour-colour diagrams with a constant reference flux K . This is convenient when tracking the changes in flux at larger disc radii (x axis, K-[24]) and in flux at smaller disc radii (y axis, K-[8]). In this paper we follow this approach, but for radiation emitted at larger disc radii in the WISE 3 and PACS 1 bands (K-W3 vs. K-P1). Furthermore, we plot the evolution in the spectral slope reference frame ($\alpha_{W3-P1}$  against $\alpha_{K-W3}$) as presented by R13 and Bustamante et al. (2015).

\section{Results}

\begin{figure}
\begin{center}
\includegraphics[width=0.47\textwidth]{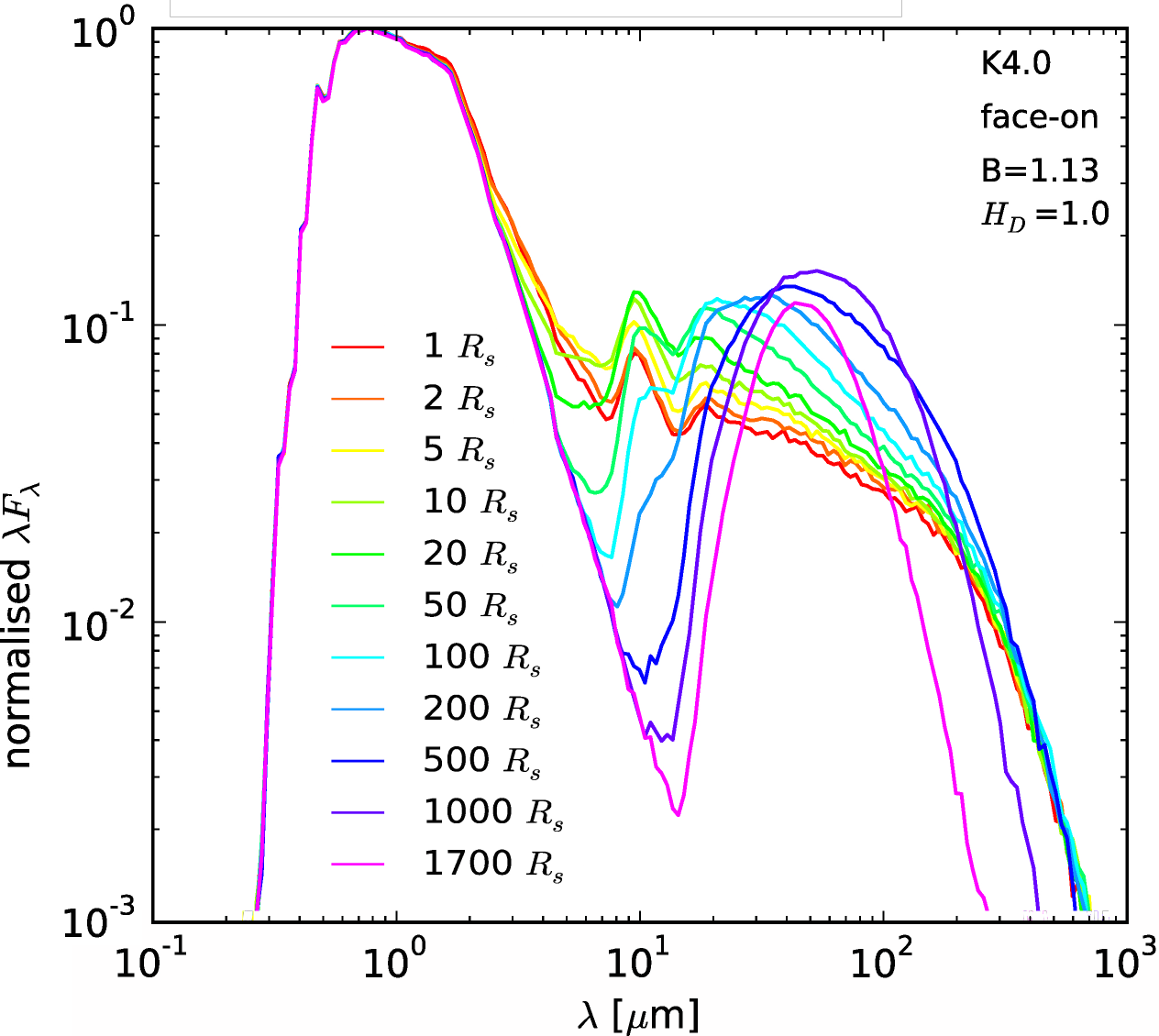}
\caption{Typical evolution of the SED for a protoplanetary disc undergoing inside-out dispersal. $R_S = 0.0573 AU$}
\label{fig:sed}
\end{center}
\end{figure}

\begin{figure*}
\begin{center}
\includegraphics[width=0.94\textwidth]{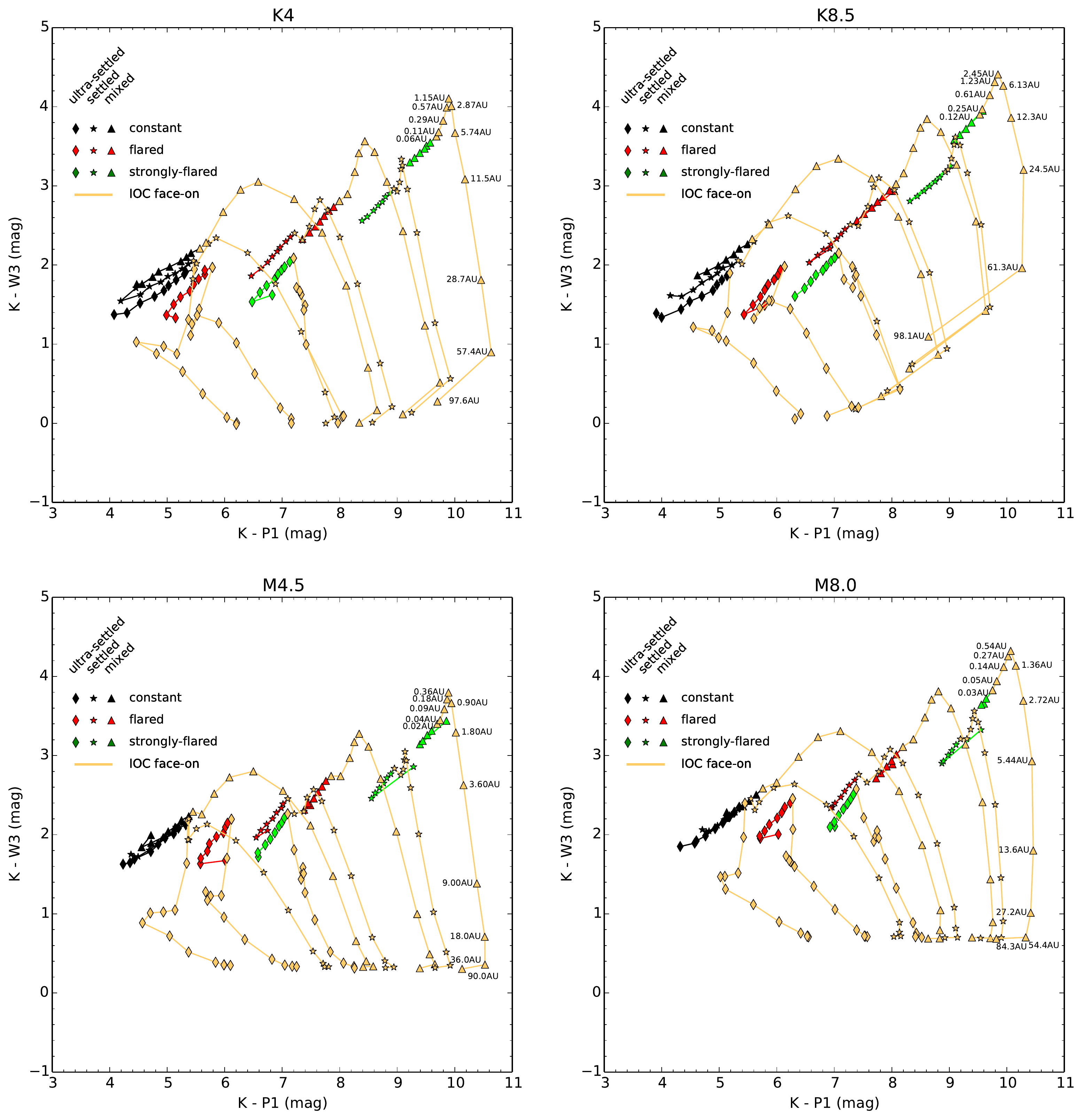}
\caption{Colour-colour evolution of primordial (black, red, green symbols) and transition discs (orange symbols and lines), for four different spectral type intervals. The orange markers are in steps of 1, 2, 5, 10, 20, 50, 100, 200, 500  times the sublimation radii, which is spectral type dependent. The K-W3 colour reddens until material at smaller radii is removed. Same goes for the K-P1 colour but the reddening in 70$\mu$m (P1) band remains dominant longer.}
\label{fig:colourcolour}
\end{center}
\end{figure*}

\begin{figure*}
\begin{center}
\includegraphics[width=0.94\textwidth]{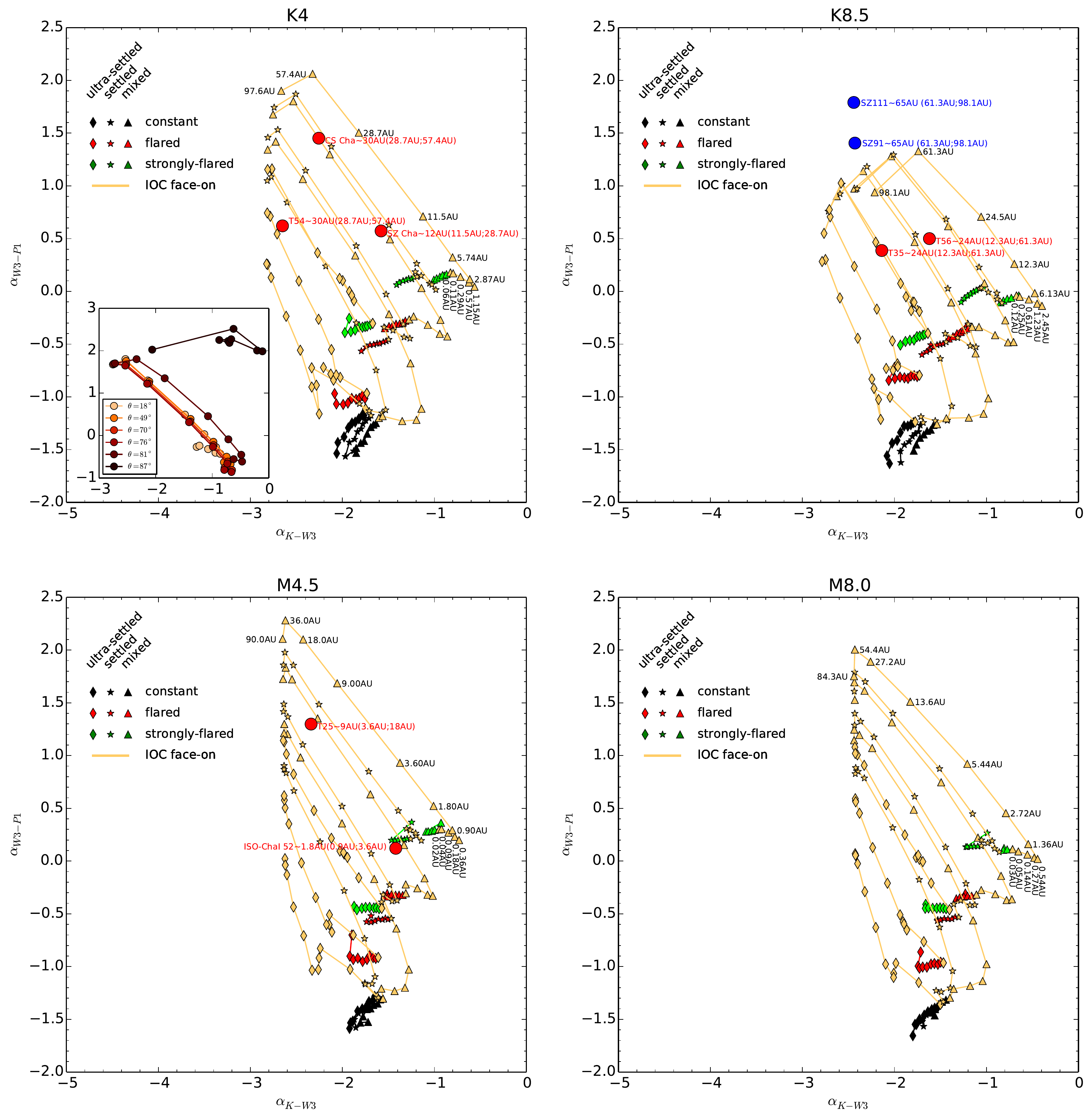}
\caption{SED slopes between the 12$\mu$m (W3) and the 70$\mu$m (P1) bands as a function of the SED slopes between the K  band and the W3 band. The models and respective symbols are the same as in Figure 2.  The dependance on inclination is shown in the small subpanel in the K4 panel for a flared mixed disc model. The measurements of R13 and B15 are included in the panels corresponding to the closest match to the spectral type of the individual objects. The inner hole for these objects was estimated roughly from our models assuming a face-on, flared, mixed disc.}
\label{fig:alpharibas}
\end{center}
\end{figure*}

\begin{figure*}
\begin{center}
\includegraphics[width=0.94\textwidth]{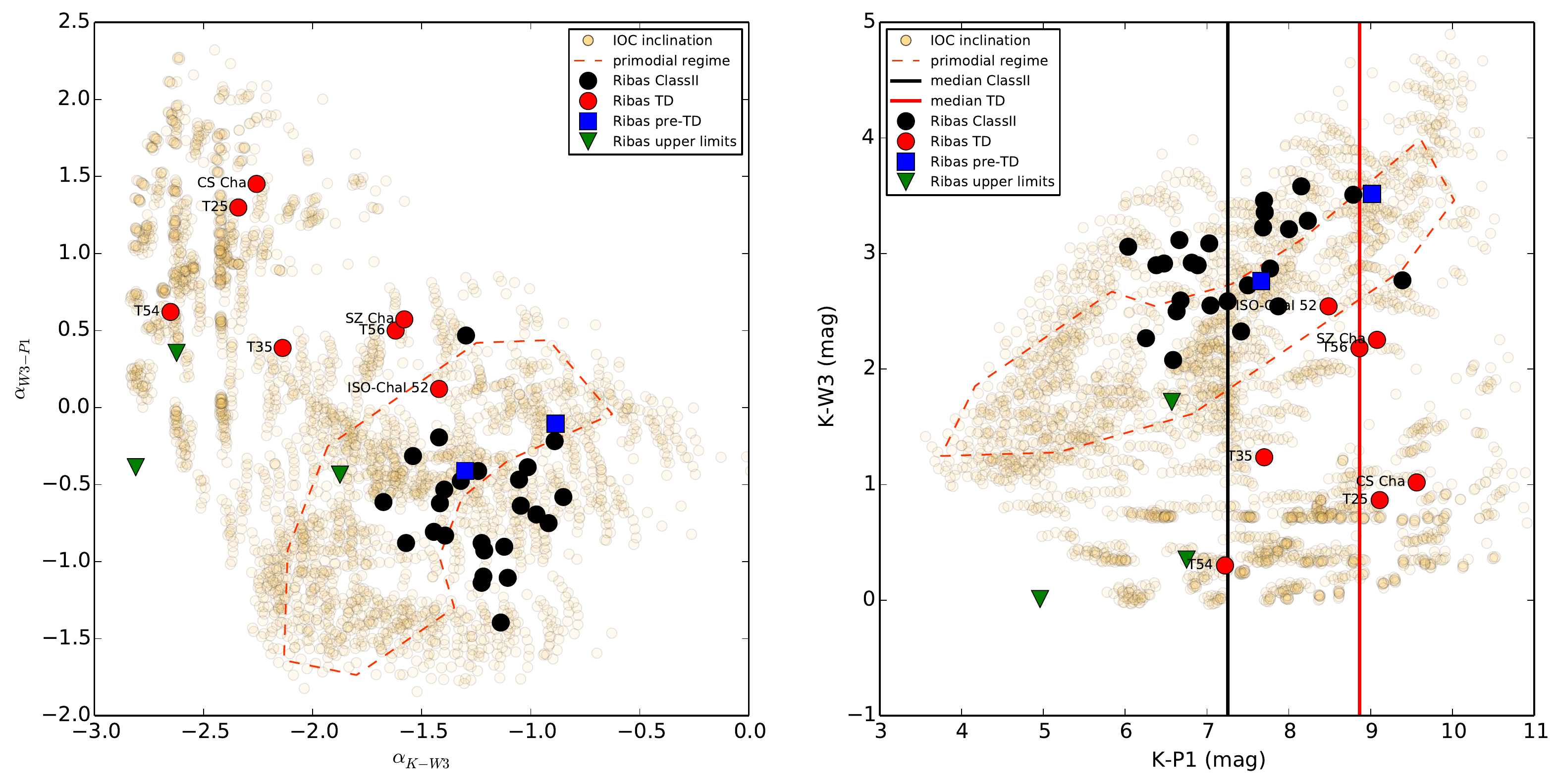}
\caption{Combination of all spectral types of Figure 3 (left) and Figure 2 (right) over-plotted with the R13 sample of primordial and transition discs. The low opacity orange circles represent the RT models for disc clearing from the inside out. The red-dashed box encloses the locus of primordial objects in the respective diagrams. The median values for transition and primordial discs from R13's sample are plotted as the vertical red and black lines, respectively. The median of the R13 TD correspond to the the inside-out clearing regime clearly showing an excess in the 70$\mu$m (P1) band.
}
\label{fig:inclination}
\end{center}
\end{figure*}

The systematic excess reported by R13 and Bustamante et al. (2015) at PACS bands in the transition discs in Chamaleon can indeed be fully understood as a natural consequence of inside-out dispersal.  To illustrate this we show in Figure$~\ref{fig:sed}$ (see also Koepferl 2012) the evolution of the spectral energy distribution of a protoplanetary disc undergoing inside-out dispersal. The disc is modelled following the methodology of Robitaille et al. (2006), where all equations and relevant parameters are described in detail. We show as an example a disc with B = 1.13, H$_{\rm D}$ = 1.0, corresponding to a well mixed flared disc, illuminated by a central star of spectral type K4 (see Table~1 of Koepferl et al. 2013 for the exact stellar atmosphere model used). The SEDs are plotted for inner holes given as multiples of the sublimation radius, which is approximately 0.06AU for a K4 star.
The figure shows that as the hotter dust from the inner radius is progressively removed the impinging stellar energy is redistributed at longer wavelengths, since radiation from the central star is reprocessed by cooler dust at larger disk radii. Both at Mid-Infrared (MIR) and FIR wavelengths the fluxes become stronger, reaching a peak when the inner hole temperature corresponds roughly to the temperature of a blackbody peaking at the given wavelength. Once the hole becomes larger and the temperature at the inner edge cooler the  fluxes become smaller again. For that reason TDs experience a brightening at MIR and then at FIR wavelengths during their inside-out evolution. Puffing up of the inner hole rim slightly dampens out this effect, but the behaviour is qualitatively the same, as shown by models in hydrostatic equilibrium, which allows for this effect.  This point was already made by Robitaille et al. (2006) in reference to the brightening of the MIR flux and applies similarly here to the FIR flux.


This behaviour can also be seen in the inside-out evolutionary tracks on a colour-colour diagram which compares a reference flux, like the unchanging K  magnitude, mostly due to stellar emission, to bands which are dominated by disc emission in the MIR or in the FIR. We show this in Figure~{\ref{fig:colourcolour} for the K-W3 versus K-P1 plane, respectively. Here we plot the evolutionary tracks for four different spectral types (K4, K8.5, M4.5 and M8.0) using a reference flux K  and probing the evolution of the outer disc with W3 and P1. 

In Figure~\ref{fig:alpharibas} we plot the SED slopes of our inside-out clearing models of face-on discs between the W3 and the P1 bands ($\alpha_{W3-P1}$) as a function of the SED slope between the K-band and W3 ($\alpha_{K-W3}$). The four panels correspond again to the four spectral types K4, K8.5, M4.5 and M8.0. The tracks are for the same models plotted in Figure~\ref{fig:colourcolour} and follow the same legend. The six transitional discs confirmed by R13 and showing excess FIR fluxes are overplotted for the spectral type that is closest to the literature value. Unfortunately there is not always an exact match in spectral type due to the course spectral type grid in the Koepferl et al. (2013) models, nevertheless the figure shows that all discs fall roughly on the inside-out clearing tracks, suggesting that the MIR-FIR flux combination of the TDs in the R13 sample is typical of objects with an inner cavity.

The tracks intervals in the Figure are for multiples of the sublimation radius and follow the same sequence as in Figure~\ref{fig:sed}, the location of the R13 sources on those tracks corresponds therefore to a given hole size, which is also indicated on these figures. The hole sizes can be therefore roughly read out from our tracks, provided the inner disc radius is high enough and tracks for a similar spectral type are available. We find the inner disc radii obtained from the simple comparison to our model tracks to be in most cases roughly consistent to the values obtained in the literature for these objects using tailored radiative transfer modeling. We describe in more details our results as follows. 

\begin{itemize}
\item We find inner disc radii of 28-58AU for CS Cha, which was classified as a K6 by Luhman (2007). This is a known binary, which may make the modeling of its circumbinary disc somewhat complicated, however the value of 40AU obtained by Espaillat et al. (2007, 2011) is consistent with our inferred range. 
\item We obtain for SZ Cha an inner radius between 12AU and 28AU, although a value closer to the lower limit of this range seems to be preferred, this is somewhat lower than the value of $\sim$30AU proposed by Kim et al. (2009). The small inconsistency is however not worrying given that we are comparing to K4 tracks while this object is classified to be a K0 (Rydrgren 1980). 
\item For T35 we obtain a value between 12 and 63~AU which compares well with the 15AU value from Espaillat et al. (2011). This star was classified as an M0 by Gauvin \& Strom (1992) and we compared its FIR colours with our K8.5 tracks. 
\item T25 (M3, Luhman et al., 2008) was modeled by Kim et al. (2009) who infered an inner radius of 8~AU, which matches the value of 4-18AU indicated by our M4.5 tracks. 
\item A tentative inner radius measurement for T56 (M0.5, Gauvin \& Strom 1992) from our models of M4.5 spectral type is approximately 24~AU, which is close to the value of 18AU inferred by Kim et al. (2009).  
\item R13 and Espaillat et al. (2011) found ISO ChaI52 to be a rather peculiar object with a very flat SED between 12 and 70 $\mu$m. In any case, ISO-ChaI52 falls in an area of our model tracks that is very degenerate, and even a rough estimate of its inner hole size is not possible from this colour combination.
\end{itemize}

We note that we obtain the closest agreement for those objects for which a close spectral type match for the model track is available.

A detailed comparison with the recent results of Bustamante et al. (2015) for TDs in the Lupus cloud is somewhat more complicated. Three of the candidate TDs in this sample, namely 2MASSJ1608149, SSTc2dJ161029.6-392215 and Sz123, lie just above the $\alpha_{W3-P1} > 0$ line, in the degenerate area of the diagrams (i.e. overlapping with primordial discs). This suggests that these objects, if they are transition discs, must have a relatively small inner cavity or perhaps an optically thick inner disc. The ambiguity of a classification based only on the W3 and P1 colour combination becomes evident for these objects, for which additional photometry measurements are needed to unveil their true nature.
In the Bustamante et al. (2015) sample, only Sz111 and Sz91 (spectral types M1.5 and 0.5, respectively, Comer\'on 2008) have large enough inner holes ($> 65AU$) to show sufficient FIR brightening to distinguish them from primordial discs. However these two objects (blue points in Figure 3) lie above the area tracked by the models for a K8.5 star, which is formally the closest match to the true spectral types in our course grid. A comparison with the M4.5 model tracks would return more sensible results, with inner hole sizes consistent with the rough observational estimate. 

The exact location of the inside-out clearing tracks on the colour-colour (or slopes) diagrams is somewhat dependant on the disc inclination, which is in most cases not known. The insert in Figure~3 shows the distribution of tracks for a typical well mixed flared disc at inclinations between 18-87 degrees, equally spaced in the cosine of the inclination angle. The well-known degeneracy between disc inclination and geometry is clearly seen. However one sees that at all but the largest inclinations the tracks converge and follow a similar behaviour, returning the same inner hole size estimate. This suggests that, while disc geometry and inclination cannot be derived from this colour combination alone, it is still possible to obtain an approximate inner radius estimate that is robust against this degeneracy, at least for inner radii larger than a few AU and inclinations lower than about 85 degrees. 

While knowledge of the spectral type of the central object and a finer spectral type grid of models are both highly desirable for the determination of the inner hole radius from the tracks, the pure identification of a large hole TD can be carried out, to a certain degree of confidence, regardless of spectral type. This is shown in Figure 4, where we plot all of our model points for the inside-out clearing tracks (orange circles) and the locus of primordial discs (red dashed line). For small inner hole sizes the loci of primordial and transition discs coincide on these diagrams. This is obvious, given that the FIR wavelengths probe material at larger disc radii. For holes larger than some tens of AUs the cutoff at $\alpha_{W3-P1} > 0$, empirically determined by R13 is indeed appropriate. One must be aware however that a census based solely on FIR detection would miss out a whole population of TDs with smaller inner hole sizes. Furthermore objects near the cutoff, run the risk of being primordial objects of a given geometry. The black and red vertical lines in Figure 4 show the median of the K-P1 colour for the TDs and primordial discs, respectively. Irrespectively of spectral type the clear shift of the TDs colour compared to the primordial discs is consistent with the expectations from an inside-out clearing evolution.

\section{Conclusions}

We have used the radiative transfer models of Koepferl et al. (2013) to track the FIR colour evolution of protoplanetary discs that are dispersing from the inside out. We have demonstrated that a brightening in the FIR is a natural consequence of this modus of evolution. 
By comparing our model tracks with Herschel 12 and 70$\mu$m photometry data (WISE 3 and PACS 1) from R13 and Bustamante et al. (2015) of transition discs in the Chamaleon I and II and Lupus star forming regions from the Herschel Gould Belt Survey (Andr\'e et al. 2010) we have shown that we the location of those objects on the colour colour plane can be explained by simple radiative transfer effects for inner cavity evolution. 

We further used our model tracks to obtain estimates of the inner disc radii for these objects, finding our values to be  reasonably robust against disc inclination and roughly consistent with previous results, which were obtained via more sophisticated, tailored radiative transfer modeling of these sources. The uncertainty on our estimates could be greatly reduced by producing more calculations with a finer parameter space grid in spectral type for the central sources and disc inner hole size. This is beyond the scope of this work and will be presented in a forthcoming paper. The new grid should provide a method to quickly extract inner radius values from far-infrared photometry data, however we caution that, while colour-colour diagrams are useful to understand the general behaviour of a population of discs, more accurate inner hole sizes are to be obtained by modeling the full SED. 

It is worth stressing at this point that a brightening at a given observed wavelength is only expected for discs with cavities whose inner radius corresponds to a temperature similar to the temperature of a blackbody peaking at the observed wavelength. Hence these models can explain also why some TDs show the brightening and some do not. 

We finally note that while a brightening of the FIR fluxes is a simple consequence of the removal of the inner disc and hence a redistribution of the impinging stellar radiation to longer wavelengths by the colder outer disc material, this effect does not explain the excess millimetre flux of a class of transition disc identified by Owen \& Clarke (2012) with large inner holes and high accretion rates. At mm wavelengths the discs become optically thin and the radiative transfer effects discussed in this paper are not relevant anymore. This is clearly seen in Figure 1, where it is shown that all SEDs converge at mm wavelengths apart from those corresponding to the very largest hole sizes, where mass removal becomes important. 

Our work thus supports the suggestion by other authors that a census of TDs using FIR colours would be useful to improve TD statistics and further our understanding of the disc dispersal mechanism. However one should also bear in mind the limitations intrinsic to using FIR photometry alone, which is only sensitive to sources with large enough inner hole sizes (some tens of AU depending on spectral type).


\section{Acknowledgements}
We thank the anonymous referee for useful comments that helped to improve our paper. 
This research made use of Astropy, a community-developed core Python package for Astronomy (Astropy 2013) and matplotlib, a python plotting library Hunter (2007).
JEO acknowledges support by NASA through Hubble
Fellowship grant HST-HF2-51346.001-A awarded by the
Space Telescope Science Institute, which is operated by
the Association of Universities for Research in Astronomy, Inc., for NASA, under contract NAS 5-26555.

\label{lastpage}

\end{document}